\def\mathrm{\rm}
\documentstyle[epsf]{l-aa}
\def\infig#1#2#3{\epsfxsize=#3cm \centering{\mbox{\epsfbox{#2}}}\vspace{-0.4cm}}

\begin{document}

   \thesaurus{06         
              (08.03.2;  
               08.06.3;  
               08.09.2)  
             }
\title{Do Si stars undergo any rotational braking?
\thanks{Based on data from the ESA Hipparcos satellite}
}
\author{P. North}

\institute{Institut d'Astronomie de l'Universit\'e de Lausanne, 
              CH-1290 Chavannes-des-bois, Switzerland
	}
\offprints{P.~North}
\date{Received December 1997 / Accepted 1998}
\maketitle

\markboth{P. North: Do Si stars undergo any rotational braking?}
{P. North: Do Si stars undergo any rotational braking?}

\begin{abstract}
The old question of rotational braking of Ap Si stars is revisited on
the empirical side, taking advantage of the recent Hipparcos results.
Field stars with various evolutionary states are
considered, and it is shown that the loose correlation between their
rotational period and their surface gravity is entirely compatible with
conservation of angular momentum. No evidence is found for any loss of
angular momentum on the Main Sequence, which confirms earlier results
based on less reliable estimates of surface gravity.

The importance of reliable, fundamental $T_{\rm eff}$ determinations
of Bp and Ap stars is emphasized.
\keywords{Stars: chemically peculiar -- Stars: fundamental parameters --
Stars: individual: HD 124224}
\end{abstract}

\section{Introduction}
It is well known that chemically peculiar stars of the Ap and Am types are
rotating slower than their normal counterparts (e.g. North 1994). The question
then arises, whether slow rotation is acquired during the main sequence life
of the star, or before its arrival on the ZAMS, i.e. during the proto-stellar
phase. Havnes \& Conti (1971) had suggested that magnetic stars undergo
magnetic braking during their main sequence lifetime, due to mass accretion
from the interstellar medium, while Strittmatter \& Norris (1971) proposed
the same, but due to mass loss. These theoretical considerations seemed to get
support from observational evidence when Wolff (1975, 1981), Stift (1976) and
Abt (1979) found some correlation between the radii or ages of Ap stars and 
their rotational period obtained from their photometric or spectroscopic 
variation. On the contrary, Hartoog (1977) concluded that magnetic Ap
stars in young clusters do not rotate faster than those in older clusters,
and this conclusion was also reached by North (1984a, b, 1985, 1986, 1987),
Borra et al. (1985) and Klochkova \& Kopylov (1985). The apparent correlation
between radius and rotational period has been commented by Hensberge et al.
(1991), who conclude that this correlation is real but possibly due to a
detection bias depending on the inclination angle, and by Stepien (1994), who
concluded on the contrary that this correlation does simply not exist, if
spurious rotational periods are duly excluded.

Using the $\log P_{\mathrm rot}$ vs. $\log g$ diagram for field stars,
North (1985, 1986, 1992) showed that for Si stars, there is indeed a trend
towards longer periods for low-gravity stars, but which can be entirely
explained by conservation of angular momentum as the star evolves with
increasing radius within the main sequence.

In this note, we revisit the $\log P_{\mathrm rot}$ vs. $\log g$ diagram for 
field stars having both a rotational period in the literature and a reliable 
surface gravity, the latter being either spectroscopic or obtained from 
Hipparcos data.

\section{The sample}
\subsection{Spectroscopic surface gravities}
The sample has been built from two parts. First was considered the list of
silicon stars for which North \& Kroll (1989, hereafter NK89) give a 
spectroscopic estimate
of $\log g$ based on the profile of the H$_\beta$ line. The estimate given
in column 5 of their Table 1 was adopted, and corrected for a constant shift
\begin{equation}
\log g(corrected) = \log g(\mathrm H_\beta) + 0.14
\end{equation}
which takes into account the systematic error displayed in their Fig. 16,
although not exactly according to their Eq. 16 which would imply too large
$\log g$ values for some stars. The intersection of this list with all stars
having a known rotational period in the literature was then done, using an
updated version of the database of Renson et al. (1991) which is a digital
version of the catalogue of Renson (1991). Two stars in the list
of NK89 which had no known period have been added, since their
period has now been determined thanks to the Hipparcos mission (Perryman et al.
1997): they are HD 154856 ($P=1.9525$~days) and HD 161841 ($P=3.21048$~days).
The original sample of NK89 was biased in favour of low
{\it photometric} gravities, hence also of low {\it spectroscopic} (and 
hopefully real) gravities, since there is a loose correlation between them.
Most stars of this sample are not very bright ($V\sim 7 - 8$), nor closeby 
enough that their parallax is significant, even for Hipparcos.

This sample contains 40 stars.

\subsection{Hipparcos surface gravities}
Second, the list of Si, SiCr or Cr stars with a Hipparcos parallax larger than 
7~mas was defined and those with a known rotational period were retained.
One star had no period in the literature but has a new one from Hipparcos
(HD 74067, $P=3.113$~days). Their mass has been interpolated in theoretical
evolutionary tracks (Schaller et al. 1992) from $T_{\mathrm eff}$ and
$M_{\rm bol}$. The effective temperature has been computed using the $X$ and $Y$
parameters of Geneva photometry calibrated by K\"unzli et al. (1997), and
corrected according to the formula $T_{\rm eff}= -230 + 0.941\times T(X,Y)$
(Hauck \& K\"unzli 1996) which replaces Eq. 1 of Hauck \& North (1993) and
where $T(X,Y)$ results from the calibration. The bolometric correction was
interpolated in Table 6 of Lanz (1984) and corrected by $\delta_{\rm BC}$
plotted in his Fig. 4a. Contrary to the previous sample, this one is not biased
regarding the distribution of the surface gravities, at least not
{\it a priori}: it is a volume-limited sample which, although surely affected
by a Malmquist-like bias, should be representative of field stars with a
more or less uniform distribution of ages. Therefore, it contains a majority
of stars which are rather close to the ZAMS in the HR diagram, just because
stellar evolution is slower there than near the core-hydrogen exhaustion phase.
If there is no {\it a priori} bias regarding the evolutionary state, one may
say, nevertheless, that the $\log g$ distribution is biased towards high values,
compared to a uniform distribution (which, then, would be strongly biased
towards large ages).

This sample contains 56 stars.

\subsubsection{Lutz-Kelker correction}
The absolute magnitudes have been corrected for the Lutz-Kelker (1973)
correction, but this correction was not applied in its original form which
assumes a constant stellar density. Indeed, the distances involved are not
negligible compared with the density scale height perpendicular to the galactic
disk, so the following generalized formulae were adopted:
\begin{eqnarray}
N(r)dr  \propto& r^2 \cos b \exp\left(-\frac{r\sin |b|}{\beta(M_V)}\right) dr \\
G(Z,\pi_{\rm o},b)=& Z^{-4}\times\exp\left(\frac{\sin |b|}{\beta(M_V)
 \pi_{\rm o}}\left[1-\frac{1}{Z}\right]\right) \nonumber \\
& \times\exp\left(-\frac{(Z-1)^2}{2(\sigma/\pi_{\rm o})^2}\right) \\
Z \equiv \frac{\pi}{\pi_{\rm o}}&
\end{eqnarray}
Let us recall that the correction on the absolute magnitude then reads:
\begin{equation}
<\Delta M(\epsilon)> = \frac{5\int_{\epsilon}^{\infty}\log Z
G(Z,\pi_{\rm o},b)dZ}
{\int_{\epsilon}^{\infty}G(Z,\pi_{\rm o},b)dZ}
\end{equation}
where $\epsilon = 0.2$, $\beta(M_V)$ is the scale height of the star density
above the galactic plane tabulated by Allen (1976), $\pi$ is the true parallax
and $\pi_{\rm o}$ is the observed parallax affected by a gaussian error
$\sigma$.

\subsubsection{Visual absorption}
The absolute magnitude also had to be corrected for the visual absorption, even
though it remains negligible in most cases. Since Cramer (1982) found that
the colour excess $E[U-B]$
defined in the Geneva system was almost the same for Bp, Ap members of clusters
as for normal B, A members -- and with a smaller dispersion than $E[B-V]$ --
this colour excess was used, corrected using Cramer's relation
$E[U-B](Bp,Ap) = E[U-B](X,Y)-0.009$ where $E[U-B](X,Y)$ is the colour excess
obtained using the intrinsic colours (of normal stars) of Cramer (1982).
$A_V$ is then obtained through $E(B-V)=1.28 E[U-B]$, since
$E[U-B] = 0.658 E[B-V]$ (Cramer 1994) and
$E(B-V) = 0.842 E[B-V]$ (Cramer 1984),
and $R = A_V/E(B-V) = 3.25 + 0.25 (B-V)_{\rm o} + 0.05 E(B-V)$ (Olson 1975).

\subsubsection{Fundamental parameters from Hipparcos parallaxes}
Once the effective temperature and bolometric correction are determined from 
photometry and the absolute magnitude from Hipparcos parallaxes as described 
above, it becomes possible to pinpoint the star on a theoretical HR diagram,
the luminosity being obtained from
\begin{equation}
\log(L/L_\odot) = -0.4 (M_V - 4.72 + B.C.)
\end{equation}
Then, we assume that Bp stars follow standard, solar-composition evolutionary
tracks (since the chemical peculiarities are limited to superficial layers
only) and the mass can be interpolated from $T_{\rm eff}$ and $\log (L/L_\odot)$
(using successive 3rd-degree splines in luminosity, $T_{\rm eff}$ and overall
metallicity $Z$, with $Z=0.018$)
whenever there is a one-to-one relation between these quantities. The latter
condition is not fulfilled near the core-hydrogen exhaustion phase, when
$T_{\rm eff}$ increases, then decreases again, and in this domain we always
assumed the star to lie on the lower, continuous branch of the evolutionary
track, which also corresponds to the slowest evolution, hence to the higher
probability. This assumption, if violated, will lead to a mass overestimate 
no larger than five percent.

The radius is directly obtained from
\begin{equation}
\log(R/R_\odot) =\frac{1}{2}\log(L/L_\odot)-2\log (T_{\rm eff}/T_{\rm eff\odot})
\end{equation}
and the surface gravity from
\begin{eqnarray}
\log g = \log\left({M \over M_\odot}\right) + 4 \log\left({T_{\rm eff} \over
T_{\rm eff\odot}}\right)\nonumber \\ - log\left({L \over L_\odot}\right) + 4.44
\end{eqnarray}
The latter equation shows how strongly $\log g$ depends on $T_{\rm eff}$, which
remains a crucial quantity. The error on it was generally assumed to be
5 percent. The errors on the other quantities are estimated using
the usual, linearized propagation formulae, but caring for the correlations
between $L$, $T_{\rm eff}$ and $M$.

The results are displayed on Table 1.
\begin{table*}
\caption{Fundamental parameters of the Si and He-weak stars
derived from the Hipparcos parallaxes. The masses were obtained by interpolation
in the evolutionary tracks of Schaller et al. (1992). Note that the errors are 
multiplied by a factor 1000 for $T_{\rm eff}$, 100 for Mass, $\log (L/L_\odot)$
and $\log g$, and 10 for $R$.
The rotational period from the literature (or from Hipparcos photometry in three
cases, see text) is given in the last column.
``LK'' means ``Lutz-Kelker correction'' and is expressed in magnitudes.}
\begin{center}
\footnotesize
\begin{tabular}{rrlllllrlll}\hline
HD&M$_{\mathrm V}$&Mass [M$_\odot$]&$\log T_{\mathrm eff}$&$\log(L/L_\odot)$&$\log g$ & R [R$_\odot$]&d [pc]
&$\sigma(\pi)/\pi$&LK [mag]& P$_{\rm rot}$ [days] \\ \hline
  4778&  1.18& 2.24$\pm$  9& 3.972$\pm$ 14& 1.51$\pm$ 7&  4.12$\pm$  9&  2.2$\pm$ 2&  93& 0.07&-0.046& 2.5616  \\
  9484&  1.00& 2.34$\pm$ 12& 3.987$\pm$ 22& 1.59$\pm$ 9&  4.12$\pm$ 13&  2.2$\pm$ 3& 128& 0.09&-0.070& 0.7 ?   \\
  9531&  0.35& 2.85$\pm$ 15& 4.039$\pm$ 20& 1.96$\pm$10&  4.19$\pm$ 13&  2.7$\pm$ 4& 126& 0.10&-0.103& 0.67    \\
  9996&  0.68& 2.47$\pm$ 15& 3.987$\pm$ 23& 1.72$\pm$12&  4.01$\pm$ 14&  2.6$\pm$ 4& 149& 0.12&-0.143&8395 (23 y)?\\
 10221& -0.28& 3.12$\pm$ 12& 4.030$\pm$ 20& 2.19$\pm$ 9&  3.95$\pm$ 11&  3.6$\pm$ 5& 141& 0.08&-0.069& 3.18      \\
 11502& -0.15& 2.87$\pm$  8& 3.989$\pm$ 18& 2.05$\pm$ 6&  3.76$\pm$  9&  3.7$\pm$ 4&  63& 0.05&-0.026& 1.60920   \\
 12767& -0.60& 3.65$\pm$ 18& 4.111$\pm$ 20& 2.39$\pm$ 9&  4.14$\pm$ 12&  3.2$\pm$ 4& 114& 0.09&-0.059& 1.9       \\
 14392&  0.26& 3.07$\pm$ 14& 4.078$\pm$ 20& 2.04$\pm$ 8&  4.29$\pm$ 11&  2.4$\pm$ 3& 112& 0.08&-0.056& 4.189     \\
 18296& -0.54& 3.32$\pm$ 15& 4.036$\pm$ 20& 2.31$\pm$11&  3.89$\pm$ 12&  4.1$\pm$ 6& 125& 0.11&-0.109& 2.8842    \\
 19832&  0.25& 3.16$\pm$ 17& 4.095$\pm$ 20& 2.04$\pm$10&  4.36$\pm$ 12&  2.3$\pm$ 3& 119& 0.10&-0.096& 0.7278972 \\
 24155&  0.18& 3.39$\pm$ 20& 4.132$\pm$ 20& 2.11$\pm$12&  4.48$\pm$ 13&  2.1$\pm$ 3& 145& 0.12&-0.138& 2.535     \\
 25267& -0.30& 3.35$\pm$ 15& 4.080$\pm$ 20& 2.26$\pm$ 7&  4.11$\pm$ 11&  3.1$\pm$ 4& 103& 0.07&-0.041& 1.210     \\
 27309&  0.34& 3.06$\pm$ 12& 4.079$\pm$ 13& 2.02$\pm$ 8&  4.32$\pm$  9&  2.4$\pm$ 3&  99& 0.07&-0.052& 1.569     \\
 29305& -0.39& 3.33$\pm$ 10& 4.064$\pm$ 14& 2.29$\pm$ 5&  4.02$\pm$  7&  3.5$\pm$ 3&  54& 0.03&-0.006& 2.94      \\
 32549& -0.98& 3.29$\pm$ 17& 3.985$\pm$ 23& 2.37$\pm$12&  3.47$\pm$ 14&  5.5$\pm$10& 131& 0.12&-0.144& 4.64      \\
 32650&  0.05& 3.08$\pm$ 13& 4.059$\pm$ 20& 2.10$\pm$ 7&  4.15$\pm$ 11&  2.9$\pm$ 4& 117& 0.06&-0.037& 2.73332   \\
 34452& -0.42& 3.95$\pm$ 21& 4.160$\pm$ 20& 2.42$\pm$10&  4.34$\pm$ 12&  2.6$\pm$ 4& 144& 0.10&-0.097& 2.4660    \\
 40312& -1.05& 3.38$\pm$  8& 3.997$\pm$ 13& 2.42$\pm$ 6&  3.49$\pm$  7&  5.5$\pm$ 5&  54& 0.04&-0.018& 3.6190    \\
 49976&  1.13& 2.21$\pm$ 11& 3.955$\pm$ 23& 1.51$\pm$ 8&  4.04$\pm$ 13&  2.3$\pm$ 3& 104& 0.08&-0.067& 2.976     \\
 54118&  0.35& 2.73$\pm$  9& 4.022$\pm$ 17& 1.89$\pm$ 5&  4.03$\pm$  9&  2.7$\pm$ 3&  87& 0.04&-0.015& 3.28      \\
 56455&  0.02& 3.25$\pm$ 14& 4.096$\pm$ 20& 2.13$\pm$ 7&  4.29$\pm$ 11&  2.5$\pm$ 3& 133& 0.07&-0.045& 1.93      \\
 72968&  1.06& 2.25$\pm$  9& 3.960$\pm$ 14& 1.55$\pm$ 7&  4.04$\pm$  9&  2.4$\pm$ 3&  84& 0.07&-0.047& 11.305    \\
 74067&  0.45& 2.57$\pm$  7& 3.988$\pm$ 13& 1.82$\pm$ 6&  3.93$\pm$  7&  2.9$\pm$ 3&  87& 0.05&-0.024& 3.11299   \\
 74521& -0.02& 3.01$\pm$ 16& 4.033$\pm$ 20& 2.11$\pm$10&  4.04$\pm$ 13&  3.3$\pm$ 5& 131& 0.11&-0.103& 7.0501    \\
 89822&  0.76& 2.57$\pm$  9& 4.025$\pm$ 16& 1.72$\pm$ 6&  4.18$\pm$  9&  2.2$\pm$ 2&  93& 0.05&-0.020& 7.5586    \\
 90044&  0.71& 2.51$\pm$ 12& 4.002$\pm$ 22& 1.73$\pm$ 8&  4.07$\pm$ 12&  2.4$\pm$ 3& 110& 0.08&-0.054& 4.379     \\
 92664& -0.37& 3.86$\pm$ 17& 4.154$\pm$ 20& 2.38$\pm$ 8&  4.35$\pm$ 11&  2.5$\pm$ 3& 146& 0.07&-0.053& 1.673     \\
103192& -0.55& 3.36$\pm$ 15& 4.044$\pm$ 20& 2.33$\pm$10&  3.90$\pm$ 12&  4.0$\pm$ 6& 117& 0.10&-0.091& 2.34      \\
112381&  1.40& 2.26$\pm$ 12& 3.999$\pm$ 24& 1.45$\pm$ 9&  4.29$\pm$ 13&  1.8$\pm$ 3& 105& 0.09&-0.076& 2.8       \\
112413&  0.24& 3.00$\pm$  9& 4.060$\pm$ 14& 2.04$\pm$ 5&  4.21$\pm$  8&  2.6$\pm$ 2&  34& 0.04&-0.011& 5.46939   \\
114365&  0.83& 2.80$\pm$ 13& 4.069$\pm$ 20& 1.81$\pm$ 8&  4.45$\pm$ 11&  1.9$\pm$ 3& 108& 0.08&-0.055& 1.27      \\
115735&  0.48& 2.55$\pm$  7& 3.990$\pm$ 13& 1.80$\pm$ 6&  3.96$\pm$  7&  2.8$\pm$ 3&  85& 0.05&-0.024& 0.77 ?    \\
116458& -0.17& 2.95$\pm$ 11& 4.012$\pm$ 22& 2.09$\pm$ 8&  3.81$\pm$ 11&  3.5$\pm$ 5& 146& 0.08&-0.063& 147.9     \\
119419&  1.09& 2.62$\pm$ 13& 4.048$\pm$ 20& 1.69$\pm$ 9&  4.45$\pm$ 12&  1.9$\pm$ 3& 116& 0.08&-0.071& 2.6006    \\
124224&  0.42& 3.03$\pm$ 17& 4.084$\pm$ 13& 1.97$\pm$ 8&  4.37$\pm$  9&  2.2$\pm$ 2&  82& 0.07&-0.046& 0.52068   \\
125248&  1.19& 2.24$\pm$  9& 3.972$\pm$ 14& 1.51$\pm$ 8&  4.12$\pm$  9&  2.2$\pm$ 3&  93& 0.08&-0.063& 9.2954    \\
125823& -1.27& 5.69$\pm$ 30& 4.248$\pm$ 20& 3.07$\pm$10&  4.20$\pm$ 12&  3.7$\pm$ 5& 134& 0.10&-0.089& 8.817744  \\
126515&  1.03& 2.29$\pm$ 17& 3.970$\pm$ 24& 1.57$\pm$15&  4.06$\pm$ 16&  2.3$\pm$ 5& 155& 0.15&-0.200& 129.95    \\
129174& -0.39& 3.49$\pm$ 14& 4.094$\pm$ 14& 2.33$\pm$ 9&  3.98$\pm$  9&  3.2$\pm$ 4& 100& 0.09&-0.067& 2.24 ?    \\
133652&  0.68& 3.05$\pm$ 14& 4.113$\pm$ 12& 1.89$\pm$ 9&  4.57$\pm$  9&  1.8$\pm$ 2&  99& 0.09&-0.082& 2.304     \\
133880&  0.09& 3.17$\pm$ 18& 4.079$\pm$ 20& 2.12$\pm$11&  4.22$\pm$ 13&  2.7$\pm$ 4& 134& 0.11&-0.116& 0.877485  \\
140728&  0.48& 2.58$\pm$  7& 3.998$\pm$ 13& 1.81$\pm$ 6&  3.99$\pm$  7&  2.7$\pm$ 2&  98& 0.05&-0.020& 1.29557   \\
142301& -0.57& 4.41$\pm$ 36& 4.193$\pm$ 20& 2.59$\pm$18&  4.35$\pm$ 17&  2.7$\pm$ 6& 161& 0.17&-0.311& 1.459     \\
142884&  0.53& 3.45$\pm$ 22& 4.160$\pm$ 20& 2.03$\pm$13&  4.67$\pm$ 14&  1.7$\pm$ 3& 133& 0.13&-0.176& 0.803     \\
149822&  0.61& 2.58$\pm$ 14& 4.010$\pm$ 22& 1.78$\pm$10&  4.07$\pm$ 13&  2.5$\pm$ 4& 140& 0.11&-0.101& 1.459     \\
152308&  0.68& 2.43$\pm$ 12& 3.976$\pm$ 22& 1.71$\pm$11&  3.97$\pm$ 13&  2.7$\pm$ 4& 146& 0.11&-0.111&1.10 (or 0.92)?\\
166469&  0.49& 2.62$\pm$ 15& 4.012$\pm$ 22& 1.81$\pm$10&  4.04$\pm$ 13&  2.6$\pm$ 4& 140& 0.10&-0.112& 2.9       \\
170000&  0.21& 2.99$\pm$ 10& 4.058$\pm$ 14& 2.03$\pm$ 6&  4.21$\pm$  8&  2.7$\pm$ 2&  89& 0.04&-0.017& 1.71649   \\
170397&  1.07& 2.35$\pm$  9& 3.993$\pm$ 13& 1.57$\pm$ 8&  4.16$\pm$  9&  2.1$\pm$ 2&  89& 0.07&-0.055& 2.1912    \\
175362& -0.52& 5.17$\pm$ 31& 4.249$\pm$ 20& 2.78$\pm$12&  4.45$\pm$ 14&  2.6$\pm$ 4& 140& 0.12&-0.152& 3.67375   \\
183806& -0.22& 2.89$\pm$ 14& 3.976$\pm$ 22& 2.07$\pm$11&  3.68$\pm$ 13&  4.1$\pm$ 7& 142& 0.16&-0.126& 2.9       \\
187474&  0.27& 2.70$\pm$ 11& 4.004$\pm$ 22& 1.90$\pm$ 9&  3.94$\pm$ 12&  2.9$\pm$ 4& 108& 0.09&-0.075& 2345      \\
199728&  0.43& 3.00$\pm$ 20& 4.078$\pm$ 20& 1.97$\pm$13&  4.35$\pm$ 14&  2.3$\pm$ 4& 143& 0.14&-0.177& 2.2       \\
203006&  0.99& 2.36$\pm$  8& 3.989$\pm$ 13& 1.60$\pm$ 6&  4.12$\pm$  8&  2.2$\pm$ 2&  58& 0.05&-0.024& 2.122     \\
221006&  0.27& 3.38$\pm$ 14& 4.135$\pm$ 20& 2.08$\pm$ 7&  4.51$\pm$ 11&  2.0$\pm$ 2& 118& 0.06&-0.033& 2.3       \\
223640&  0.08& 3.21$\pm$ 15& 4.089$\pm$ 13& 2.12$\pm$10&  4.27$\pm$ 10&  2.5$\pm$ 3& 103& 0.10&-0.090& 3.735239  \\
\hline
\end{tabular}
\end{center}
\end{table*}

\subsection{Comparison between different sources of $\log g$}
A comparison between photometric and spectroscopic $\log g$ values was already
shown by NK89. Fig. 1 shows how photometric and Hipparcos values compare,
for Si and HgMn stars lying closer than 100~pc to the Sun. The diagrams
look exactly the same as in the comparison of photometric vs. spectroscopic
values, i.e. the Si stars are strongly scattered ($\sigma_{\rm res}=0.273$~dex)
while the HgMn stars follow the one-to-one relation much more closely
($\sigma_{\rm res}=0.080$~dex), with the exception of HD 129174, a visual double
which was excluded from the fit.
\begin{figure}[th!]
\infig{8.4}{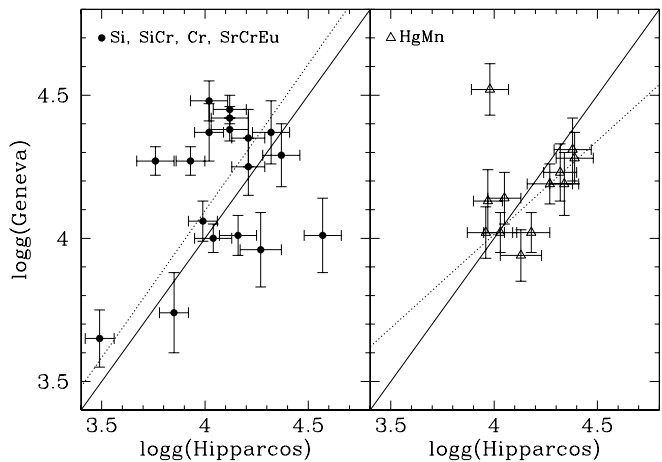}{8.4}
\caption{Comparison between photometric and Hipparcos $\log g$ values for Si
(left, full dots) and HgMn (right, open triangles) stars closer than 100~pc.
The continuous line is the one-to-one relationship, while the dotted line is a 
least-squares fit which takes into account similar errors on both axes. The 
discrepant point on the right panel is HD 129174, a visual double excluded from 
the fit.}
\end{figure}
The nice behaviour of the HgMn stars in this diagram inspires confidence in the
value of Hipparcos gravities.

The comparison between spectroscopic and Hipparcos gravities for Si stars is
shown in Fig. 2, where all stars of the list of NK89 having Hipparcos parallaxes
with $\sigma(\pi)/\pi\leq 0.14$ are plotted (please note that some of them do
not appear in Table 1 because they have $\pi < 7$~mas). Unfortunately, only six
objects fulfill this criterion; among them, four are on the equality line
within the errors (at least within $2 \sigma$), while two are clearly below.
\begin{figure}[th!]
\infig{8.8}{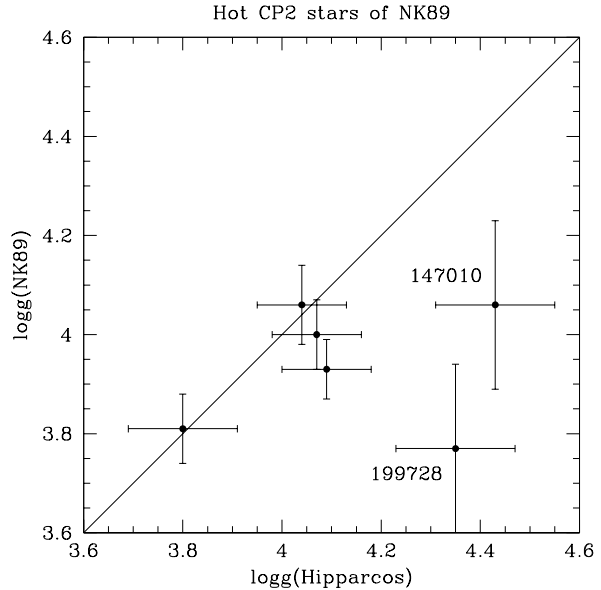}{8.8}
\caption{Comparison between spectroscopic and Hipparcos $\log g$ values for Si
stars with $\sigma(\pi)/\pi\leq 0.14$. The continuous line is the one-to-one 
relationship.}
\end{figure}
The two outsiders are HD 147010 and HD 199728. Interestingly, these stars have
the largest photometric amplitude, as shown in Table 2 where the peak-to-peak
amplitude in Str\"omgren's $u$ band (or Geneva $[U]$ band) is given with its
source. This suggests that photometry overestimates $T_{\rm eff}$ in cases of
extreme peculiarities\footnote{Interestingly, Abt \& Morrell (1995)
classify HD 199728 as F0:Vp while it is surely hotter than 10000~K, even
though photometry tends to overestimate its effective temperature.},
and is quite coherent with the fact that, in Table 1,
some stars have $\log g$ values (determined from Hipparcos luminosities)
around 4.5, which is about 0.2 dex more than the theoretical ZAMS value.
This is probably due to an overestimate of their effective temperature.
It seems that those Ap stars having a more
or less fundamental $T_{\rm eff}$ value have on average less extreme
peculiarities than those having a good rotational period (hence a large
photometric amplitude) and considered here, so that the photometric calibration
tends to overestimate $T_{\rm eff}$ for some of the latter. Nevertheless,
no systematic correction will be made on $\log g$ in this sample, because
the bias strongly depends on the individual stars.
\begin{table}
\caption{Peak-to-peak amplitudes of the 6 stars having both a spectroscopic
and a Hipparcos $\log g$ value.}
\begin{center}
\begin{tabular}{rcl}\hline
\multicolumn{1}{c}{HD}&$u$ or $[U]$& Source \\
& total ampl.& \\ \hline
49976 & 0.055 & Catalano \& Leone (1994) \\
90044 & 0.060 & Manfroid \& Renson (1994) \\
94660 & 0.035 & Hensberge (1993) \\
147010& 0.080 & North (1984c) \\
164258& 0.016 & Catalano \& Leone (1994) \\
199728& 0.127 & Renson (1978) \\
\hline
\end{tabular}
\end{center}
\end{table}
\subsection{Hipparcos radii versus $v\sin i$}
In order to test the validity of the radii obtained using Hipparcos parallaxes,
a comparison between the observed projected rotational velocities and equatorial
velocities obtained from the formula of the oblique rotator model
\begin{equation}
V_{eq} [km\,s^{-1}] = 50.6\times R [R_\odot]/P [days]
\end{equation}
is shown on Fig. 3. The sources of $v\sin i$ are Abt \& Morrell (1995), Levato
et al. (1996), Renson (1991) and Uesugi \& Fukuda (1981).
\begin{figure}[th!]
\infig{8.8}{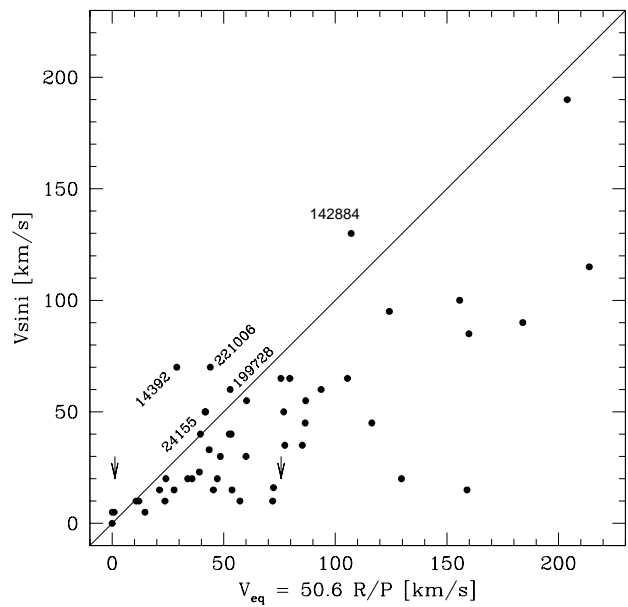}{8.8}
\caption{Comparison between the observed $v\sin i$ and the equatorial velocity
computed from the period and from the Hipparcos radius. The continuous line is 
the one-to-one relationship. Stars lying above this line are labeled by their
HD number. Arrows indicate cases where only an upper limit to $v\sin i$ is
known.}
\end{figure}
Most stars fall below the equality line, as expected from $\sin i \leq 1$;
therefore, the test appears rather successful, statistically speaking.
However, seven of them are above, at least two of which simply because of the
uncertainty on the $v\sin i$ determination (HD 126515 and HD 187474, with
$V_{\rm eq}\sim 0$). HD 199728 is only slightly above, but this may well be due
to an underestimate of its radius linked with an overestimate of its effective
temperature (see Subsection 2.3 and Fig. 2 above). This is also the case of
HD 24155, HD 142884 and HD 221006 (which have $\log g = 4.48$, 4.67 and 4.51
respectively, suggesting an overestimated radius), although the radius of the
latter star would need to be strongly underestimated. The star HD 14392 (and
possibly HD 221006 too) lies so high above the equality line that its rotational
period may be questioned. Indeed, Pyper \& Adelman (1985) proposed a period of
1.3102 days (following Winzer 1974), instead of 4.189 days proposed later by 
e.g. Adelman \& Knox (1994). The photometric curves of HD 14392 are so scattered
that the shorter period may be the right one after all; magnetic and 
spectroscopic observations should be done to settle the matter. The rotational
period of HD 221006 has been found to lie around 2.31 days by Renson (1978) and
this was confirmed by Manfroid \& Mathys (1985) and by Leone et al. (1995).
There seems to be no reason to question this value; therefore, we are left with
two possibilities: either $v\sin i=69$~km\,s$^{-1}$ (Uesugi \& Fukuda 1970) is
overestimated, or the radius is underestimated by more than 30 percent. This
appears doubtful, since $T_{\rm eff}= 13275$~K has been estimated in a 
quasi-fundamental way (with the IR Flux Method) by M\'egessier (1988) and is
only 370~K lower than our photometric estimate: such a difference does not imply
an increase of $R$ by more than 3 percent.

Finally, HD 142884 has a reliable period and its radius must be underestimated
by about 20 percent, as suggested by its very large $\log g$ (4.67). An
independant estimate of its $T_{\rm eff}$ would be extremely welcome.

\section{The $\log P_{\mathrm rot}$ vs. $\log g$ diagram}
Fig. 4 shows the distribution of stars according to their rotational period
and surface gravity. There is of course an intrinsic scatter, but on average
the width of the period distribution is relatively narrow and there are clearly
longer periods among the more evolved stars. Stars with both a small $\log g$
and a very short period are lacking. There are two stars falling below the
lower envelope in a significant way: HD~115599 and HD~150035. HD~115599 was
measured photometrically by Moffat (1977) only once a night near culmination,
so that the published period might very well be an alias of the real one.
The photometric measurements of HD 150035 made by Borra et al. (1985) do not
seem very precise, judging from the low S/N lightcurve they published.
The period of this star appears to remain highly uncertain.

It is interesting to consider the case of CU Vir or HD 124224, because in the
literature a very small $\log g$ is sometimes quoted: for instance, Hiesberger
et al. (1995) quote values as small as 3.45 to 3.60 (obtained from
spectrophotometric scans), but also 4.2 and 3.71. The latter two values come
from the same $uvby\beta$ photometric indices but through two different
calibrations. The Hipparcos data, together with $T_{\rm eff} = 12130$~K
obtained from Geneva photometry, point to $\log g = 4.37\pm 0.09$, i.e. the
star is very close to the ZAMS. If a higher effective temperature is adopted,
like $T_{\rm eff} = 13000$~K, the result becomes worse, with
$\log g = 4.50\pm 0.08$ (the error on $\log g$ was computed assuming an error
of only 400~K on $T_{\rm eff}$). The conclusion that CU Vir is unevolved seems
unescapable and is coherent with the fact that no Bp or Ap star has a rotational
period significantly shorter than 0.5 days (the record is held by HD 60431, with
$P_{\rm rot} = 0.47552$~days, see North et al. 1988). This may bear some
importance in view of the fact that CU Vir is the only Ap star for which a
period change has been unambiguously identified (Pyper et al. 1998). Any
explanation for this intriguing discovery will have to take into account the
unevolved state of the star.

The full and broken lines drawn in Fig. 4 are kinds of evolutionary tracks: 
assuming an initial period of 0.5~days (respectively 4.0~days), they show how a 
star rotating as a rigid body will evolve, if no loss of angular momentum 
occurs. These lines essentially reflect how the moment of inertia changes with
evolution for stars having 2.5 and 5~M$_\odot$. They depend in a negligible way
on the mass and are entirely compatible with the observations. They were
established starting from the conservation of angular momentum:
\begin{equation}
I\omega = I_0\omega_0
\end{equation}
where $\omega$ is the star's angular velocity, $I$ the moment of inertia and
the subscript $0$ indicates initial value (i.e. on the ZAMS). For the period,
one has
\begin{equation}
P = \frac{2\pi}{\omega}\, \Rightarrow \,P = P_0 \frac{I}{I_0}
\, \Rightarrow \, \log P = \log P_0 + \log \frac{I}{I_0}
\end{equation}
How the moment of inertia changes with evolution is provided by the models
of Schaller et al. (1992), through a code kindly provided by Dr. Georges
Meynet.

The two steep, straight dotted lines illustrate the extreme case of
conservation of angular momentum in concentric shells which would rotate
rigidly but glide one over the other without any viscosity, i.e. without
the least radial exchange of angular momentum. In such a case, the moment of
inertia of each shell of mass $\delta m$ and radius $r$ reads
\begin{equation}
I = \frac{2}{3} \delta m r^2
\end{equation}
and in particular, the outermost shell having $r=R$ and being the only one 
observed, one gets
\begin{eqnarray}
P = P_0 \left(\frac{R}{R_0}\right)^2 = P_0 \frac{g_0}{g} \\
\log P = \log P_0 + \log g_0 -\log g
\end{eqnarray}
Surely this case is an ideal and not very realistic one, but it is shown for
illustrative purpose.

Do Si stars undergo any rotational braking during their life on the Main
Sequence? Because of the decreasing number of stars with decreasing $\log g$,
the statistics remains a bit small, and doubling the number of stars in the
range $\log g < 3.8$ would be very useful. Nevertheless, the data are
entirely compatible with nothing more than
conservation of angular momentum for a rigidly rotating star. They may be
marginally consistent with the dotted lines whose slope is 1 (conservation of
angular momentum for independent spherical shells): if these lines are
interpreted as betraying some {\it loss} of angular momentum through some
braking mechanism yet to be understood, then this loss cannot increase the
period by more than about
\begin{equation}
\log P = \log P_0 + 0.325 (\log g_0 -\log g)
\end{equation}
meaning a relative increase of no more than 82 percent during the whole
Main Sequence lifetime. This is only a fraction of the increase due to
angular momentum conservation alone (for a rigid sphere).

The whole reasoning has been applied to a mix of stars with various masses
(between 2.2 and 5.7~M$_\odot$), but if any magnetic breaking exists, its
efficiency might well be a sensitive function of mass. Then, one would need
a larger sample, allowing $\log P$ vs $\log g$ diagrams to be built
separately for stars in narrow mass ranges. The sample as a whole would
not need to be enlarged in an unrealistic way: it is especially the evolved
stars which are crucial for the test, so increasing their number from
13 (for $\log g < 3.8$) to about 50 or 70 would probably be enough
to answer the question on firmer
grounds. Spectroscopic observations would be needed to estimate $\log g$
(and hopefully $T_{\rm eff}$!) and photometric ones to determine the
periods.

\begin{figure}[th!]
\infig{8.8}{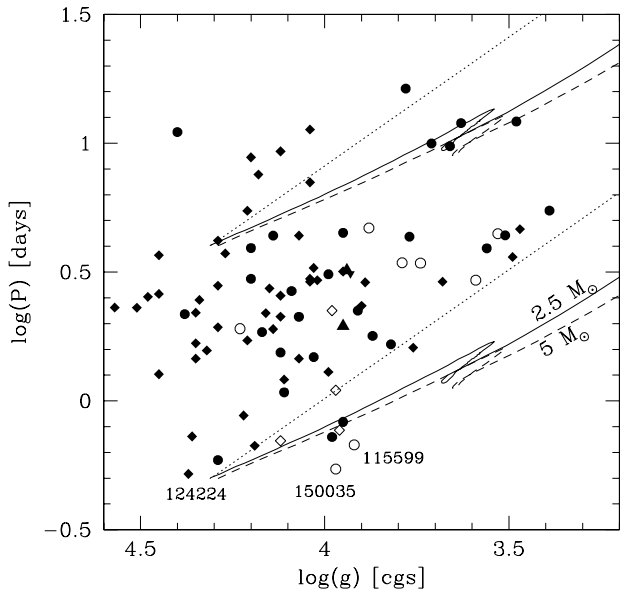}{8.8}
\caption{Rotational period versus surface gravity. Full symbols represent stars
with a reliable period, open symbols are for possibly ambiguous periods. Round
dots (and triangles) represent stars with a spectroscopic value of $\log g$,
while diamonds are for stars with $\log g$ determined from Hipparcos data.
The three triangles are for stars with a rotational period newly determined
from Hipparcos photometry (the upside-down triangle has $\log g$ determined
from Hipparcos, the others from spectroscopy). The continuous and broken lines 
represent the evolution of the period predicted from that of the moment of 
inertia, under the assumption of rigid-body rotation and for initial periods of 
0.5 and 4 days. The dotted lines show the ideal case of conservation of angular
momentum in independent spherical shells.}
\end{figure}

\section{Conclusion}
New surface gravities of magnetic Bp and Ap stars obtained from the Hipparcos
parallaxes, as well as homogeneous spectroscopic gravities, have been used to
reconsider how the rotational period of such stars varies with age.
The result is entirely consistent with previous
works suggesting that field Si stars do not undergo any significant magnetic 
braking during their life on the Main Sequence; it is also more firmly based
than earlier studies made on field Ap stars. Therefore, the slow rotation
of these objects must be a property acquired {\it before} they arrive on the
ZAMS. How this occurs has just been explored by Stepien (1998) but further 
investigations remain worthwhile.

On the other hand, this study has shown that $\log g$ values obtained
from Hipparcos luminosities may be overestimated by up to 0.2 dex for some
extreme Ap stars, probably through an overestimate of their $T_{\rm eff}$. This
shows how badly fundamental determinations of $T_{\rm eff}$ are needed for
these stars.

\acknowledgements{This work was supported in part by the Swiss National
Fondation for Scientific Research. I thank Fabien Carrier for his update of
the database of Ap stars, which was widely used, and Dr. Georges Meynet for
the code which allowed me to compute stellar moments of inertia.
Daniel Erspamer computed the visual absorptions and Dr. Laurent Eyer
provided a list of new periods of Ap stars obtained with Hipparcos.
This research has made use of the Simbad database, operated at CDS, Strasbourg, 
France. I thank the referee for his constructive criticism.}


\begin{thebibliography}{}
\bibitem{} Abt H.A., 1979, ApJ 230, 485
\bibitem{} Abt H.A., Morrell N., 1995, ApJS 99, 135
\bibitem{} Adelman S.J., Knox J.R.Jr., 1994, A\&AS 103, 1
\bibitem{} Allen C.W., 1976, Astrophysical Quantities, 3rd edition, Athlone
Press, p. 250
\bibitem{} Borra E.F., Beaulieu A., Brousseau D., Shelton I., 1985, A\&A 149, 
266
\bibitem{} Catalano F.A., Leone F., 1994, A\&AS 108, 595
\bibitem{} Cramer N., 1982, A\&A 112, 330
\bibitem{} Cramer N., 1984, A\&A 132, 283
\bibitem{} Cramer N., 1994, thesis No 2692, Geneva University
\bibitem{} Hartoog M.R., 1977, ApJ 212, 723
\bibitem{} Hauck B., North P., 1993, A\&A 269, 403
\bibitem{} Hauck B., K\"unzli M., 1996, Baltic Astronomy 5, 303
\bibitem{} Havnes O., Conti P.S., 1971, A\&A 14, 1
\bibitem{} Hensberge H., 1993, in: Peculiar versus normal phenomena in A-type
and related stars, IAU Coll. No 138, eds. M.M. Dworetsky, F. Castelli and
R. Faraggiana, A.S.P. Conf. Series Vol. 44, p. 547
\bibitem{} Hensberge H., Van Rensbergen W., Blomme R., 1991, A\&A 249, 401
\bibitem{} Hiesberger F., Piskunov N., Bonsack W.K., Weiss W.W., Ryabchikova T.,
Kuschnig R., 1995, A\&A 296, 473
\bibitem{} Klochkova V.G., Kopylov I.M., 1985, Soviet Astronomy 29, 549
\bibitem{} K\"unzli M., North P., Kurucz R.L., Nicolet B., 1997, A\&AS 122, 51
\bibitem{} Lanz T., 1984, A\&A 139, 161
\bibitem{} Leone F., Catalano F.A., Manfr\`e M., 1995, A\&A 294, 223
\bibitem{} Levato H., Morrell N., Solivella G., Grosso M., 1996, A\&AS 118, 231
\bibitem{} Lutz T.E., Kelker D.H., 1973, PASP 85, 573
\bibitem{} Manfroid J., Renson P., 1994, A\&A 281, 73
\bibitem{} Manfroid J., Mathys G., 1985, A\&AS 59, 429
\bibitem{} M\'egessier C., 1988, A\&AS 72, 551
\bibitem{} Moffat A.F.J., 1977, IBVS 1265
\bibitem{} North P., 1984a, Thesis No 2117, Geneva University
\bibitem{} North P., 1984b, A\&A 141, 328
\bibitem{} North P., 1984c, A\&AS 55, 259
\bibitem{} North P., 1985, A\&A 148, 165
\bibitem{} North P., 1986, in: Upper main sequence stars with anomalous
abundances, eds. C.R. Cowley, M.M. Dworetsky and C. M\'egessier, Reidel, p. 167
\bibitem{} North P., 1987, A\&AS 69, 371
\bibitem{} North P., 1992, in: Stellar magnetism, eds. Yu.V. Glagolevskij and
I.I. Romanyuk, Sankt-Petersburg, p. 72
\bibitem{} North P., 1994, in 25th Meeting of the European Working Group on CP
stars, eds. I. Jankovicz and I. Vincze, Szombathely, p. 3
\bibitem{} North P., Babel J., Lanz T., 1988, IBVS 3155
\bibitem{} North P., Kroll R., 1989, A\&AS 78, 325
\bibitem{} Olson B.I., 1975, PASP 87, 349
\bibitem{} Perryman M.A.C., H{\o}g E., Kovalevsky J., Lindegren L.,
Turon C., 1997, The Hipparcos and Tycho Catalogues, ESA SP-1200
\bibitem{} Pyper D.M., Adelman S.J., 1985, A\&AS 59, 369
\bibitem{} Pyper D.M., Ryabchikova T., Malanushenko V., Kuschnig R.,
Plachinda S., Savanov I., 1998, A\&A (submitted)
\bibitem{} Renson P., 1978, IBVS 1391
\bibitem{} Renson P., 1991, Catalogue G\'en\'eral des Etoiles Ap et Am, Institut
d'Astrophysique - Universit\'e de Li\`ege
\bibitem{} Renson P., Kobi D., North P., 1991, A\&AS 89, 61
\bibitem{} Schaller G., Schaerer D., Maeder A., Meynet G., 1992, A\&AS 96, 269
\bibitem{} Stepien K., 1994, in: Chemically peculiar \& magnetic stars, eds.
J. Zverko and J. Ziznovsky, Tatranska Lomnica, p. 8
\bibitem{} Stepien K., 1998, in: Proceedings of the 26th Workshop of the
European Working Group on Chemically Peculiar Stars held in Vienna on
27-29 October 1997, eds. J. \v{Zi}\v{z}\v{n}ovsk\'{y}, P. North and A. Schnell, 
Tatranska Lomnica (in press)
\bibitem{} Stift M.J., 1976, A\&A 50, 125
\bibitem{} Strittmatter P.A., Norris J., 1971, A\&A 15, 239
\bibitem{} Uesugi A., Fukuda I., 1970, Contr. Kwasan Obs. Univ. Kyoto No. 180
\bibitem{} Uesugi A., Fukuda I., 1981, Revised Catalogue of Stellar Rotational
Velocities, CDS, Strasbourg, France
\bibitem{} Winzer J.E., 1974, Thesis, University of Toronto
\bibitem{} Wolff S.C., 1975, ApJ 202, 121
\bibitem{} Wolff S.C., 1981, ApJ 244, 221
\end{thebibliography}
\end{document}